\documentclass[aps,twocolumn,showpacs,preprintnumbers,amsmath,amssymb,nofootinbib,superscriptaddress,showkeys]{revtex4-1}

\usepackage{amsmath,epsfig}
\usepackage{longtable}
\allowdisplaybreaks[4] 

\usepackage[normalem]{ulem}
\usepackage{color}

\newcommand{\bear}{\begin{eqnarray}}
\newcommand{\eear}{\end{eqnarray}}

\newcommand{\ba}{\begin{array}{c}}
\newcommand{\ea}{\end{array}}
\newcommand{\nn}{\nonumber}

\newcommand{\be}{\begin{equation}}
\newcommand{\ee}{\end{equation}}

\newcommand{\Frac}[2]{\frac{\displaystyle #1}{\displaystyle #2}}

\newcommand{\cO}{{\cal O}}

\newcommand{\gsim}{\stackrel{>}{_\sim}}
\begin{document}



\preprint{ FTUAM-13-14}
\preprint{IFT-UAM/CSIC-13-069}

\title{Pad\'e Approximants and Resonance Poles}

\author{Pere Masjuan} \email{masjuan@kph.uni-mainz.de}
 \affiliation{Institut f\"ur Kernphysik, Johannes Gutenberg-Universit\"at, D-55099 Mainz,
 Germany }

\author{Juan Jos\'e Sanz-Cillero} \email{juanj.sanz@uam.es}
\affiliation{Departamento de F\'isica Te\'orica and Instituto de F\'isica Te\'orica, IFT-UAM/CSIC
       Universidad Aut\'onoma de Madrid, Cantoblanco, Madrid, Spain}

\date{\today}


\begin{abstract}

Based on the mathematically well defined Pad\'e Theory, a theoretically safe new procedure
for the extraction of the pole mass and width of a resonance is proposed. In particular,
thanks to the Montessus de Ballore theorem we are able to unfold the Second Riemann Sheet
of an amplitude to search for the position of the resonant pole in the complex plane. The method
is systematic and provides a model-independent treatment of the prediction and the corresponding
errors of the approximation.

\end{abstract}



\maketitle


\setcounter{footnote}{0}


\section{Introduction}

The rigorous quantum-mechanical definition of a resonance with given quantum numbers corresponds
to a pole in the Second Riemann Sheet (2RS) in the (analytically continued) partial-wave amplitude
of the considered scattering channel \cite{Martin:102663}.
This definition becomes independent of the background, whereas the corresponding residue provides
the probability to produce that resonance in the given process.

However, although quoting the complex pole and residue of a resonance would
be superior and highly desirable, for practical reasons this is not what one typically finds
in the PDG~\cite{PDG2012} with very few exceptions. Instead, several definitions besides the pole position in the 2RS are employed, such as a pole in the $K$-matrix, the Breit-Wigner resonance, the location of a maximum in the speed plot, the time delay, etc (see, e.g., \cite{Suzuki:2008rp,Workman:2008iv}). 

Complex energies cannot be measured and an analytic continuation to the complex plane is required. If the amplitude on the real axis is just approximated or a model, the analytic continuation might amplify the uncertainty. A model-independent procedure to explore the second Riemman sheet would then be very welcome.

The non-perturbative regime of QCD is characterized by the presence of physical resonances,
complex poles of the amplitude in the transferred energy at higher complex Riemann Sheets (instead of the physical one). In many cases, from the experimental point of view, one can obtain information about
the spectral function of the amplitude through the time-like region ($q^2 > 0$)
and also about its low-energy region through the experimental data on the space-like region ($q^2 < 0$).

In this article we develop a theoretically safe new procedure for the extraction of the pole mass, width and residue of resonances based on the mathematically well defined Pad\'e Theory~\cite{Baker}. This theory explores the features of convergence of a sequence of rational functions, called Pad\'e Approximants (PA), to the function one wants to investigate. In this regard, Pad\'e Theory provides with a set of theorems of convergence that allows us not only to propose a model-independent method for extracting resonance properties for such function but also provide a criterion for the evaluation of the error on the extraction of such resonance parameters. In particular, thanks to the Montessus de Ballore's theorem~\cite{Montessus} we are able to unfold the 2RS of a physical amplitude to search for the position of its resonant pole (if any) in the complex plane. 

When applied to physical amplitudes, PA are usually centered at the origin of energies $q^2=0$. In Refs.~\cite{Peris-VFF,Masjuan:2012wy}, the particular cases of the $\pi\pi$-Vector Form Factor (VFF) and the $\pi\gamma$-Transition Form Factor (TFF), respectively, were analyzed within the Pad\'e Theory with the main purpose of studying their low-energy behavior using the available space-like data. In particular, the first and the second derivatives of the VFF~\cite{Peris-VFF} and the TFF~\cite{Masjuan:2012wy} were precisely determined at $q^2 = 0$ trough a fit procedure to that data.

\begin{figure*}[ht]
 \includegraphics[width=6.9in]{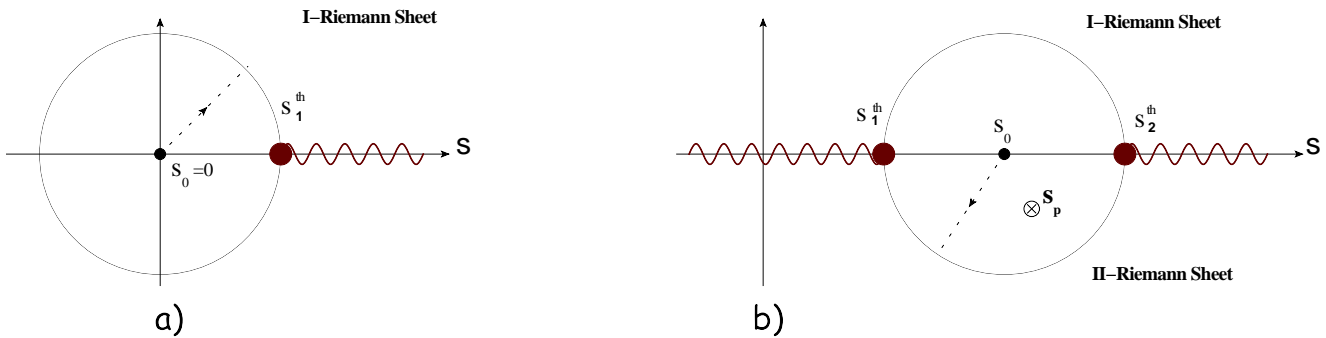}\\
\caption{Unfolding the Second Riemman Sheet with PAs centered above the branch-cut singularity. $s_1^{th}$ and $s_2^{th}$ are the production thresholds and $s_p$ is the position of the eventual resonance pole of the amplitude.}
 \label{fig:Unfold}
\end{figure*}

Despite the nice convergence and the systematical treatment of the errors obtained in such Refs.~\cite{Peris-VFF,Masjuan:2012wy}, that procedure does not allow us to obtain properties of the amplitude above the threshold,
such as in the case of the $\pi\pi$-VFF, the $\rho$-meson pole position. The reason is simple:
the convergence of a sequence of PA centered at the origin of energies ($q^2 = 0$) is limited by
the presence of the $\pi\pi$ production brunch cut, see Fig.~\ref{fig:Unfold}, panel {\it a}, with $s_1^{th}$ the production threshold. The PA sequence converges everywhere except on the cut, so the
2RS cannot be approach. Still, the mathematical Pad\'e Theory allows us to produce a model
independent determination of resonance poles when certain conditions are fulfilled. The most important one is
to center our PA sequence above the branch-cut singularity (beyond the first production threshold)
instead of at origin of energies ($q^2 = 0$), see Fig.~\ref{fig:Unfold}, panel {\it b}. This small modification also provides the opportunity to use time-like data in our study instead of the space-like one. The relevance of this model-independent method to extract resonance poles is clear since it does not depend on a particular Lagrangian realization or modelization
on how to extrapolate from the data on the real energy axis into the complex plane.
 An analogous attempt to extract the value of complex poles has been performed in Ref.~\cite{Svarc:2012pt}, based
on a Pietarinen expansion~\cite{Pietarinen:1972nk}.

Although we apply the Pad\'e method in the particular case of a physical amplitude to extract the position of a resonance pole, it is clear that it can be applied in a broader number of cases since only relies on a mathematical theory and not on a particular physical situation. We illustrate that method using a couple of examples where the properties above appear naturally.

The article is organized as follows: in Section~\ref{sec:general} we provide the main features of PA and Montessus' theorem. In Section~\ref{sec:examples} the features of that theorem are investigated in a set of analytical models where all the properties of the theorem appear naturally. Two different ways are explored: the first one, which we call ``genuine PA'', consists of constructing the approximants from the derivatives of the function around some particular energy point; and the second, which we call ``fit PA'', where the derivatives of such function are obtained from the PA after fitting them to a set of data. In Section~\ref{sec:pheno}, we give two phenomenological examples to illustrate the simplicity of the
proposed ``genuine PA'' and ``fit PA'' methods in realistic scenarios. Section~\ref{sec:comparison} compares our method of extracting resonance poles with the commonly used procedure based on the Breit-Wigner distribution model. We show that the Breit-Wigner model is equivalent to the first element on the PA sequence, and so it can easily be improved upon with our Pad\'e method. We will conclude the article in Section~\ref{sec:conclusions} pointing out  other observables which
can be analyzed in a similar way.

\section{Pad\'e Approximants}\label{sec:general}

As stated in the Introduction, the method we want to present in this article is based on the mathematical theory of Pad\'e Approximants. In the following we introduce the main features of such theory, its motivation for studying resonance poles and Montessus' theorem. Further details, demonstration of theorems and extensions can be found in the book of Baker and Grave-Morris~\cite{Baker}.

Let us consider  a function $F(x)$,  analytical in a disk $B_{\delta}(x_0)$. Then, the Taylor expansion
\begin{equation}\label{taylor}
{\cal P}_N(x,x_0)=\sum_{n=0}^{N} a_n (x-x_0)^n\,  ,
\end{equation}
\noindent
converges to $F(x)$ in $B_{\delta}(x_0)$ for $N\rightarrow \infty$, with derivatives given by $a_n=F^{(n)}(x_0)/n!$. In that situation, one usually use experimental data to extract the derivatives of $F(x)$ by polynomial fits at higher and higher order $N$. Since the experimental data have errors, one normally finds that polynomials with order higher than $\tilde{N}$ do not produce new information, with the new coefficients of order $\tilde{N}+1$ been compatible with zero. In that situation, one stops the fit procedure at order $\tilde{N}$ and takes it as one's best estimate.

The scenario changes, however, when the function $F(x)$ is not analytical anymore, for example when it has a single pole at $x=x_p$ inside the disk $B_{\delta}(x_0)$. In this case, the Taylor series does not converge any more, so we need a different procedure to extract information about the function and its derivatives.

An easy way to go beyond the range of applicability of the Taylor expansion is by invoking the so-called Pad\'e Approximants (PA) to the function $F(x)$, denoted by $P^M_N(x,x_0)$.  PA are defined~\cite{Baker} as a ratio of two polynomials $Q_M(x,x_0)$ and $R_N(x,x_0)$\footnote{$R_N(0)=1$, without loss of generality.}, of order $M$ and $N$ (respectively) in the variable $x$, with a \textit{contact} of order $M+N$ with the expansion of $F(x)$ around $x=x_0$. Thus, when expanding $P^M_N(x,x_0)$ around $x=x_0$, one reproduces exactly the first $M+N+1$ coefficients of the expansion for $F(x)$:

\begin{equation}\label{PAdef}
P^M_N(x,x_0)\,\,\,=\,\,\, F(x)\,\,\, +\,\,\, {\cal O}\bigg((x-x_0)^{M+N+1}\bigg)\,.
\end{equation}

Although polynomial fitting is more common, in general, rational approximants (i.e., ratios of two polynomials) are able to approximate the original function in a much broader range in momentum than a polynomial \cite{Baker}. This will be the great advantage of the PAs compared to other methods: they allow the inclusion of low and intermediate energy information in a rather simple way which, furthermore, can in principle be systematically improved upon~\cite{Peris-VFF,Masjuan:2012wy}. In certain cases, like when the form factor obeys a dispersion relation given in terms of a positive definite spectral function (i.e., becomes a Stieltjes function), it is known that the Pad\'e sequence is convergent everywhere on the complex plane, except on the physical cut \cite{Baker,Peris:2006ds,IAM-critic,Masjuan:2009wy}. Another case of particular interest is in the limit of QCD with an infinite number of colors, in which form factors become meromorphic functions. In this case there is also a theorem which guarantees convergence of the Pad\'e sequence everywhere in a compact region of the complex plane, except perhaps at a finite number of points (which include the poles in the spectrum contained in that region) \cite{Pommerenke,Masjuan:2007ay,Masjuan:2008fr}. In the real world, in which a general form factor has a complicated analytic structure with a cut, and whose spectral function is not positive definite, we do not know of any mathematical result ensuring the convergence of a Pad\'e sequence \cite{IAM-critic}. One just has to try the approximation on the data to learn what happens.

A special case of interest for the present work is Montessus de Ballore's theorem \cite{Montessus,SanzCilleroProc,MasjuanProc}. Montessus' theorem states that when the amplitude $F(x)$ is analytical 
inside the disk $B_\delta(x_0)$ except for a single pole at $x=x_p$ the sequence of one-pole Pad\'{e} Approximants $P_1^N(x,x_0)$ around $x_0$,

\begin{equation}\label{PAeq}
P_1^N(x,x_0)=\sum_{k=0}^{N-1}a_k(x-x_0)^k+\frac{a_N(x-x_0)^N}{1-\frac{a_{N+1} }{a_N} (x-x_0)}\, ,
\end{equation}
\noindent
converges to $F(x)$ in any compact subset of the disk excluding the pole $x_p$, i.e,

\begin{equation}\label{th}
\lim_{N\rightarrow \infty} P_1^N (x,x_0)\, =\, F(x)\, .
\end{equation}

As an extra consequence of this theorem, one finds that the PA pole  $x_{PA}=x_0+\frac{a_N}{a_{N+1}}$ converges to $x_p$ for $N\rightarrow \infty$. In the same way, the PA residue $Z_{PA}=\, - (a_N)^{N+2}/(a_{N+1})^{N+1}$ also converges. Since experiments provide us with values of $F_j$ at different $x_j$
 instead of the derivatives of our function, we can use the rational
 functions $P_1^N(x,x_0)$ as fitting functions to the data. In this way, as $N$
 grows $P_1^N(x,x_0)$ gives us an estimation of the series of derivatives and both the $x_p$
 pole position and $r_p$ residue.
 
 In this article, we restrict ourselves to the application of Montesus' theorem to the simplest case of physical amplitudes with a single-resonance pole inside the disk $B_{\delta}(x_0)$, where single-resonance pole functions demand single-pole PAs. Montessus' theorem goes, however, beyond that simplest scenario and ensures the convergence of the $P^N_M(x,x_0)$ sequence given that the disk $B_{\delta}(x_0)$ contains $M$ poles. Two resonance poles would then demand a $P^N_2(x,x_0)$ sequence, and both resonance poles will be convergently located in a systematic way. Of course, in this case, a $P^N_1(x,x_0)$ sequence will also converge (yielding the position of the resonance pole closer to $x_0$) but only up to the position of the second resonance pole. In a scenario with resonance poles but no brunch cuts, the range of convergence of the $P^N_M(x,x_0)$ sequence is defined by the position of the $M+1$ pole. In the scenario with a single-resonance pole and branch cuts that we are considering here, the $P^N_1(x,x_0)$ sequence converges as stated by the Montesus' theorem and the $P^N_2(x,x_0)$ sequence should also converge taking into account the feasibility of the second PA pole to emulate the brunch cut~\cite{IAM-critic}. Discussions along these lines are postponed to be given elsewhere.
 
Usually, as we have already said, PA are constructed around the low-energy point $x_0=0$ where $x$ is the total energy squared. For a physical amplitude, the function $F(x)$ (without a left-hand cut)~\footnote{Notice that, however, scattering amplitude partial-wave projections in general generate a left-hand cut.}) is analytic from $x=-\infty$ up to the first production threshold $x_{th}$ and within the disk $B_{x_{th}}(0)$. In the $\pi \pi$ vector form-factor case, the threshold is found to be at $x_{th}=4m_{\pi}^2$, where $m_{\pi}$ is the mass of the pion. Experiments provide $F^{exp}(x)$ at $x<0$ and can be used to extract the derivatives of the form-factor at the origin~\cite{Peris-VFF,Masjuan:2012wy}. One may also consider time-like $F^{exp}(x+i0^+)$ data at $x>x_{th}$, but PAs centered at $x_0=0$ cannot be applied to them due to the presence of the essential singularity at $x=x_{th}$~\cite{Masjuan:2007ay,Peris-VFF,Masjuan:2012wy}. However, one can still use PAs in a safe way by using Montessus' theorem, i.e., by centering them at $x_0+i0^+$ over the brunch cut between the first and the second production threshold ($x_1^{th}<x_0<x_2^{th}$). In the $\pi\pi$-VFF that would correspond to the range between pion-production threshold and the kaon one, $x_1^{th}=4m_{\pi}^2$ and $x_2^{th}=4m_K^2$ (assuming a negligible contribution from  multipion channels). In such a way we unfold the 2RS due to the analytical extension of the function $F(x)$ from the first Riemann sheet at $x+i0^+$ into the second one.

In the case of resonant amplitudes, a single pole appears in the neighborhood of the real $x$ axis in the 2RS which can be related to a hadronic state, a resonance. Once we have unfold the 2RS by locating our approximants over the brunch cut, the application of the Montessus' theorem in the disk of convergence (i.e., in the region defined between the thresholds) is straightforward and allows us to locate the position of the resonant pole if it lies inside that region. If that is the case, our $P_1^N(x,x_0)$ sequence systematically determines its position in a model independent way.

\section{Analytical examples}\label{sec:examples}

\subsection{Analytical models}

To illustrate the possibilities of our method,
we consider three different $\rho$-like models of the $\pi\pi$-VFF,
with a single pole in the 2RS at
$s_p=(0.77-i0.15/2)^2$GeV$^2$ and a logarithmic branch cut
(starting at $s=0$ for simplicity). The considered models will be of the form
\bear
\label{eq:model}
F(s) &=& \Frac{M^2}{M^2-s \, +\, \Frac{1}{\pi} M G\,  \xi(s)}\, ,
\eear
being each model specified by
\begin{eqnarray}\label{eq.models}
\mbox{\bf Model A):}& \quad &
\xi(s)\,=\,
\ln{\Frac{-s}{M^2}}  \, ,
\nn\\
\mbox{\bf Model B):}&
 &
\xi(s)\,=\, \Frac{  s}{M}
\ln{\Frac{-s}{M^2}}   \, ,
\\
\mbox{\bf Model C):}&
&
\xi(s)\,=\, \Frac{s}{M^2} \ln{\Frac{-s}{M^2}}+
\nn\\
&&
+ \frac{1}{2}\Frac{s\rho_K^2(s)}{M^2}
\left[2-\rho_K(s)\ln\left(\frac{\rho_K(s)+1}{\rho_K(s)-1}\right)\right] \, , \nn\\
\nn
\end{eqnarray}
with $M$ and $G$ conveniently tuned in each case to produce the pole at $s=s_p$.
In the last model, we
also incorporate an upper  second threshold at
$s=4m_K^2$~\cite{GomezDumm:2000fz,Oller:2000ug,SanzCillero:2002bs,SanzCillero:2009pt,Gounaris:1968mw},
with $\rho_K(s)=\sqrt{1- 4 m_K^2/s}$. We will set the upper threshold at  $4 m_K^2=1$~GeV$^2$
and take {\it model C} as our most refined one and closer to the situations one may find in real physics, where higher thresholds are also present.

There are two different ways to explore the method with these models, which we will study
in the next two subsections: the {\it genuine} PA and the {\it fitting} PA approach.

\subsection{Genuine PA}

\begin{figure}
  \includegraphics[width=3in]{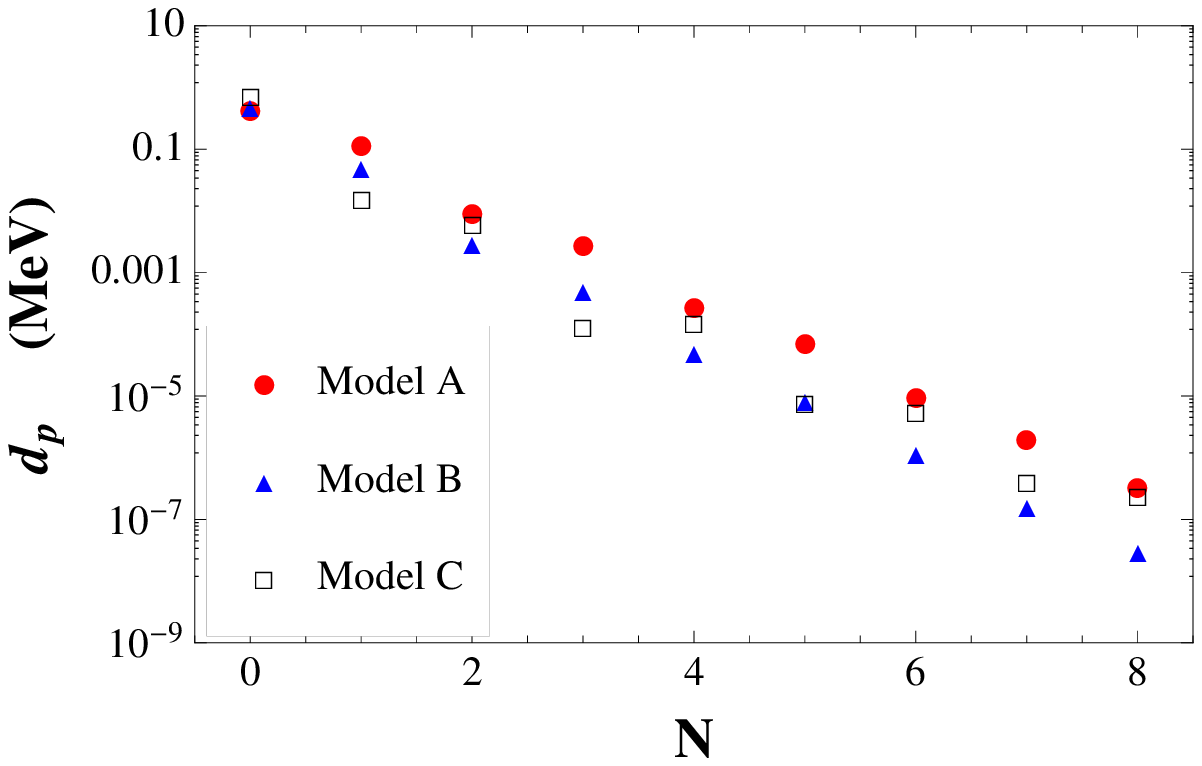}
  \includegraphics[width=3in]{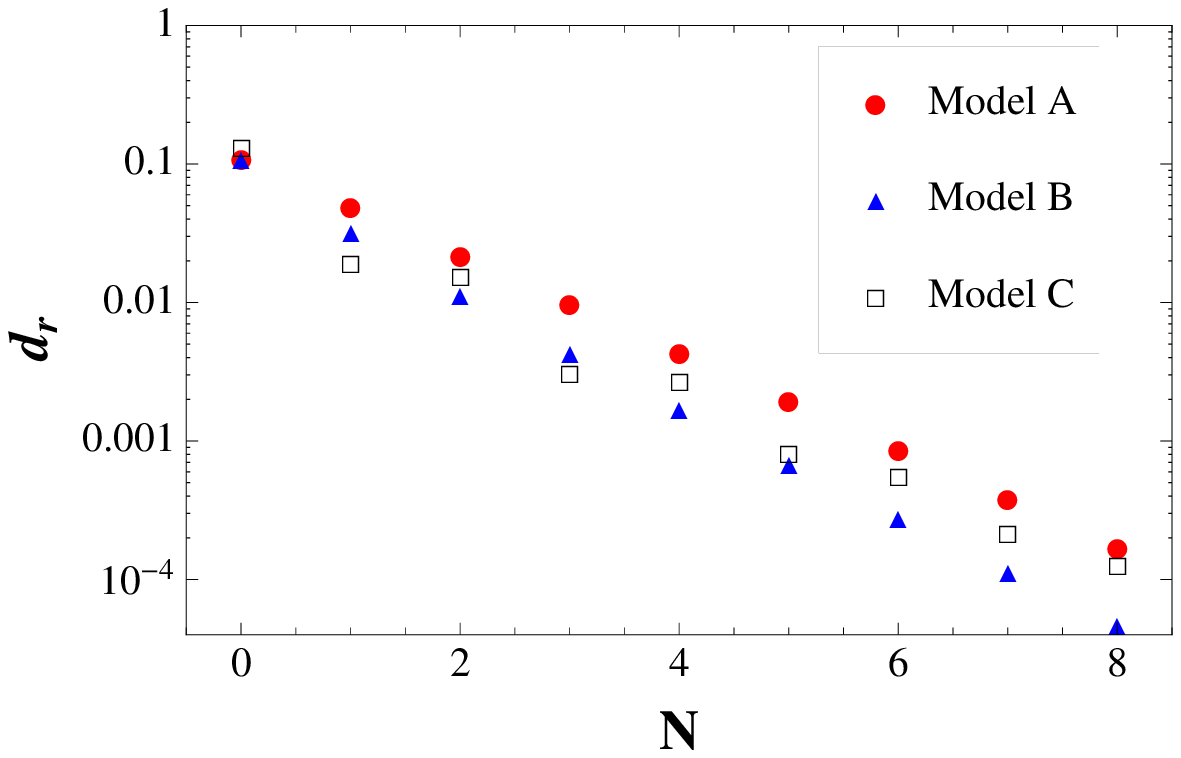}
  \\
  \caption{Rate of convergence of the $P^N_1(s,s_0)$ Pad\'{e} Approximants sequence
  corresponding to the A, B, C models in Eq.~(\ref{eq.models}). {\bf Top:} Deviation from the
  true pole position in MeV (Eq.~(\ref{eq.dpole})).
{\bf  Bottom:} relative value of the deviations in the value of the residue (Eq.~(\ref{eq.dres})).
  For all the cases we used the PA center $s_0=0.6$~GeV$^2$. }
  \label{fig:modelsABC}
\end{figure}

The first method we want to present consists on using the derivatives of $F(s)$, Eq.~(\ref{eq:model}),  around some point $s_0$ to construct
a $P_1^N(s,s_0)$ sequence and extract from them the (convergence
sequence for the) position of the pole.  The pole $s_{PA}^{(N)}$ and residue $Z_{PA}^{(N)}$ of the
$P^N_1(s,s_0)$ around a point $s_0$  are given by the analytical expressions
\bear\label{eq:poleres}
s_{PA}^{(N)}
&=& s_0 \,+\,    \Frac{a_N}{a_{N+1}} \,  ,
\nn\\
Z_{PA}^{(N)} &=& \, -\, \Frac{ (a_N)^{N+2}}{(a_{N+1})^{N+1}}
 \, ,
\eear
with $a_N=\frac{1}{N!}\frac{d^N F}{ds^N}\big|_{s=s_0}$.

In order to be able to appreciate the relevance of our approximation,
we define the distance between the predicted pole and the real pole as
\begin{equation}
d_p=\sqrt{(M_{PA}-M_{pole})^2+(\Gamma_{PA}-\Gamma_{pole})^2}\, ,
\label{eq.dpole}
\end{equation}
with $s_{PA}=(m_{PA}- i\Gamma_{PA}/2)^2$, and $s_p=(M_{pole} - i\Gamma_{pole}/2)^2$.   
This parameter $d_p$ helps us to see the rate of convergence of our sequence
of approximants for each A, B, C models in Eq.~(\ref{eq.models})
as it is illustrated in Fig.~\ref{fig:modelsABC} (PA centered at $s_0=0.6$~GeV$^2$).
In particular, the first prediction, using  $P_1^0(s,s_0)$,
has an error $d_p<1$~MeV, the second one ($P_1^1(s,s_0)$)  $d_p<0.1$~MeV,
and so on. The reader should take into account that the figure is
in logarithm scale.

It is also interesting to study the deviation of the predicted residue $Z_{PA}$ with respect to that
in the original model ($Z$):
\begin{equation}
d_r\, =\, \bigg|\Frac{Z_{PA}}{Z} \, - \, 1\bigg| \, ,
\label{eq.dres}
\end{equation}
where a very high precision is immediately obtained in the first PA orders.

\begin{figure}
  \includegraphics[width=3in]{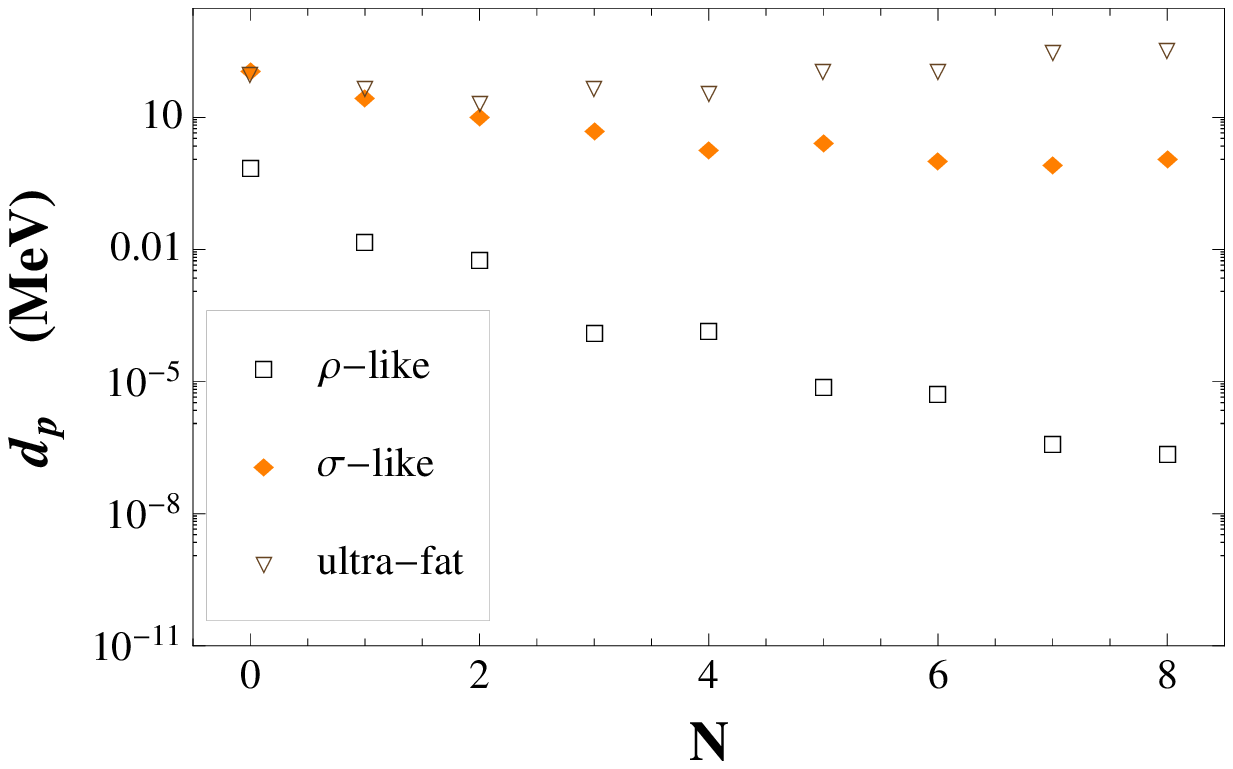}
  \includegraphics[width=3in]{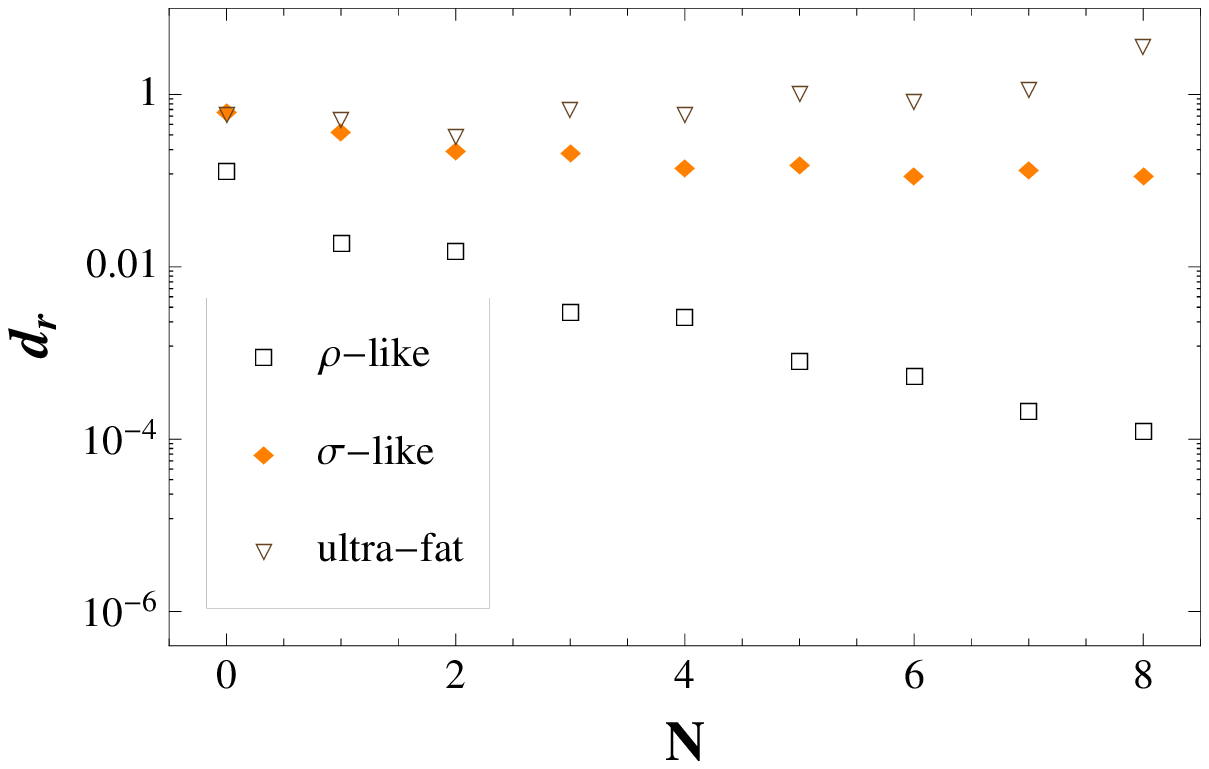}\\
  \caption{{\small
  Rate of convergence of the $P^N_1(s,s_0)$ Pad\'{e} Approximants
  sequence corresponding to the $\rho$-like,  $\sigma$-like   and
  {\it ultra-fat} versions of the {\it model C} in Eq.~(\ref{eq.models}). {\bf Top:} deviation~(\ref{eq.dpole}) of the pole position in MeV. {\bf Bottom:} relative deviation~(\ref{eq.dres}) of the residue prediction.
  }}
  \label{fig.rho+sigma+uf}
\end{figure}

At this point, a word of caution is needed. In order to successfully recover  the position
of a resonant pole using the Montessus' theorem the resonance pole we are looking for
must lie within the disk around $s_0$  where the theorem can be applied.
In our case, the disk is limited by the first production thresholds immediately below and above $s_0$;
beyond these branch-cut singularity points the convergence disk can be extended no longer.
In that sense,
the prediction for  the pole position in a $\rho$-like model should converge very fast
(where  $s_p^{\rho}=(0.77-i0.15/2)^2$GeV$^2$),
in a $\sigma$-like case (broader than a $\rho$-like particle,
with $s_p^{\sigma}=(0.5-i0.5/2)^2$GeV$^2$),
the convergence should be  slower and for an {\it ultra-fat}  particle
(e.g. $s_p^{uf}=(1-i/2)^2$GeV$^2$) there should be no  convergence at all.
This is shown for {\it model C} in Fig.~\ref{fig.rho+sigma+uf} for the PA center $s_0=0.6$~GeV$^2$.
The needed {\it model-C} parameters were
$(M,G)^{(\rho)}=(0.786,0.146)$~GeV, $(M,G)^{(\sigma)}=(0.983,1.852)$~GeV,
$(M,G)^{({\rm ultra-fat} )}=(1.846,2.915)$~GeV, resp.
One can easily observe that as the resonance pole is placed deeper and deeper into the complex plane,
the convergence of the PA in the $s$-variable  turns worse and worse and, eventually,
the sequence diverges, as one can see in the {\it ultra-fat} example.

However, the PA does not need to be constructed just in terms of the $s$--variable.
One may consider  an analytical  coordinate transformation that may improve
(or even ensure) the convergenge of the PA. In the case of two-body processes,
one of the typical kinematical variables one may take for the PA is the center-of-mass three-momentum
\bear
k&=&\Frac{ \lambda(s,m_1,m_2)^\frac{1}{2}}{2\sqrt{s}}\, ,
\eear
with $m_1$ and $m_2$ the masses of the lower threshold
two particles and the kinematic function $\lambda(x,y,z)=x^2+y^2+z^2-2x-2 x z-2 yz$.
In our models, with a massless lower threshold,
this corresponds to simply $k=\sqrt{s}/2$.  In general this transformation folds the complex plane
in such a way that the pole approaches to the real axes.
Another usual alternative is given by the conformal mapping
\bear\label{eqw}
w&=&\Frac{\sqrt{s-s_1^{th}}-\sqrt{s_2^{th}-s}}{\sqrt{s-s_1^{th}}+\sqrt{s_2^{th}-s}}\, ,
\eear
with $s_1^{th}$ and $s_2^{th}$ the position of the lower and higher thresholds, respectively.
This maps the $1^{st}$ Riemann Sheet with Im$[s]>0$  and the $2^{nd}$ Riemann Sheet with Im$[s]<0$
into the circular region  with $|w|<1$.
The  points $s=(s_1^{th},s_2^{th},\pm \infty + i\epsilon,\pm \infty - i\epsilon)$ are transformed into
into $w=(-1,1,+i,-i)$. The cut discontinuity $s\in (s_2^{th},+\infty)$ above the second threshold becomes the
Re$[w]>0$ part of the circle $|w|=1$ and the discontinuity $s\in (-\infty,s_1^{th})$
turns into its Re$[w]<0$ part. Notice that this change of variable is by no means similar to the one used in Ref.~\cite{Cherry:2000ut},  which has a completely different analytical structure. On the contrary to what happens with $w$ in Eq.~(\ref{eqw}), the $z$ variable used in~\cite{Cherry:2000ut} places the branch cuts very close to the data between the first two thresholds.    

\begin{figure}
  \includegraphics[width=3in]{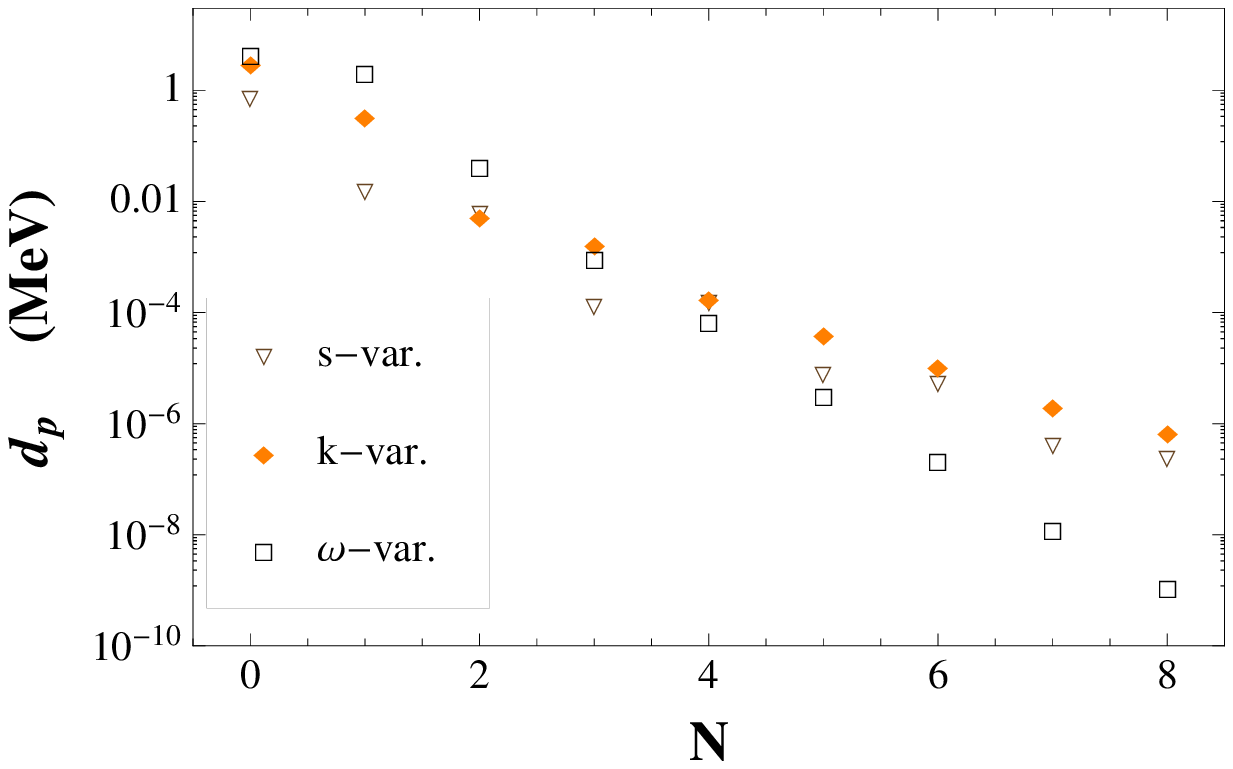}
  \includegraphics[width=3in]{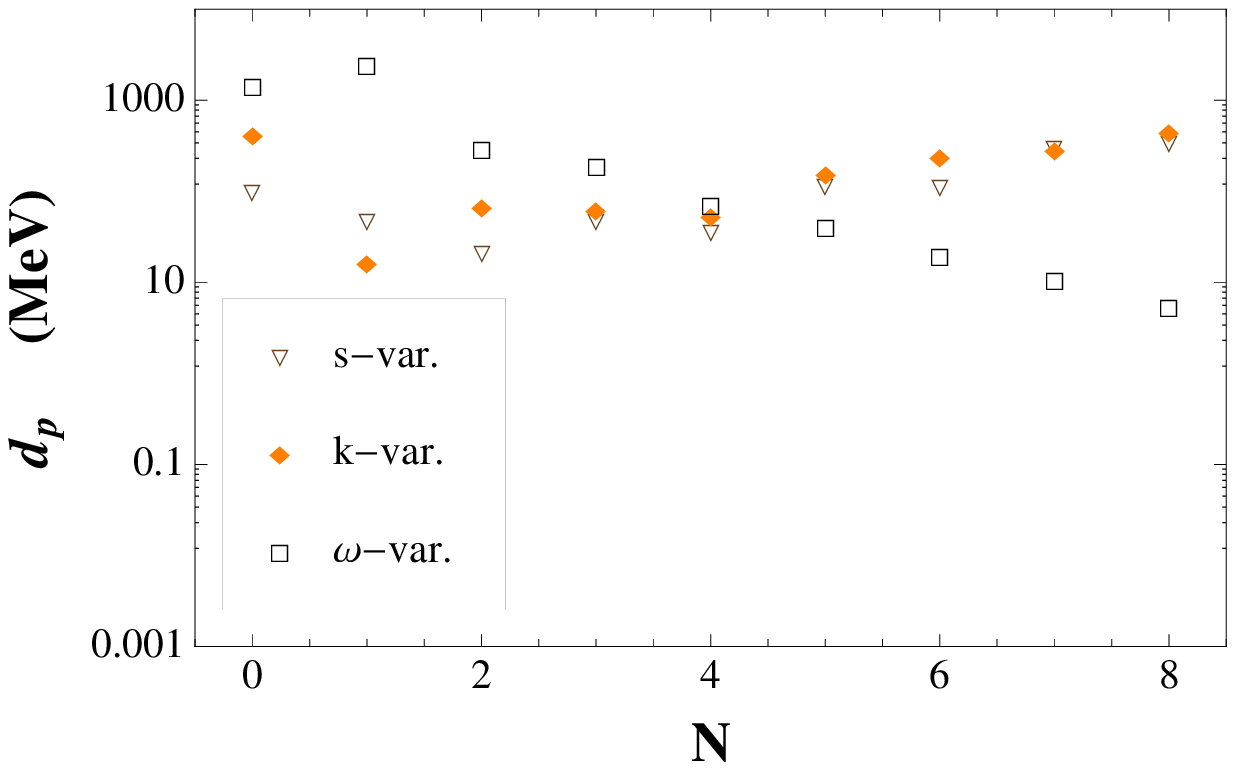}\\
  \caption{{\small
  Rate of convergence of the pole position for  the $P^N_1(s,s_0)$ Pad\'{e} Approximant
  sequence corresponding to the $\rho$-like (upper panel) and
  {\it ultra-fat}  (lower panel) resonance profiles for {\it model C} in Eq.~(\ref{eq.models}) in different variables: $s$, $k$ and $w$ (inverse empty triangles, orange rhombus, and empty squares respectively). A similar behaviour is found for the residue prediction.
  }}
  \label{fig.s-k-w}
\end{figure}

Thanks to the conformal transformation in Eq.~(\ref{eqw}), it does not matter how far away into the complex plane
the pole is located in terms of the $s$-variable; it will be always within the $|w|=1$  circle,
the Montessus' theorem will be always applicable and we will find a convergent sequence
of $P^N_1(w,w_0)$ approximants.  We can observe this feature in Fig.~\ref{fig.s-k-w}, where we compare the
pole-converge rate for both $\rho$--like (upper panel) and {\it ultra-fat} resonances (lower panel)  in terms of the
$s$, $k$ and $w$ variables. We take the {\it model C}
and the PA center $s_0=0.6$~GeV$^2$ and, correspondingly, $k_0=k(s_0)$
and $w_0=w(s_0)$.  Although, in this example, the conformal variable seems to yield a worse approximation
for low order PA, eventually it provides a  convergence  behaviour better than that for $s$ and $k$.
Indeed, in the {\it ultra-fat} resonance case where the $s$ and $k$ sequences diverge (as expected),
one can observe a slow but clear convergence in the $w$ variable.
We want to remark that
even if some variable transformations may ensure the convergence, this does not necessarily tells us
how soon or how fast this happens.

\subsection{Systematic error criterium}

Since the convergence of our PA sequence is granted by the Montesus' theorem for $N\rightarrow \infty$, we can establish  a systematical error  in our PA approach for a finite value of $N$. As the order of the approximant increases
the $s_{PA}^{(N)}$ predictions converge to  the actual pole $s_p$ of the amplitude $F(s)$.
The step will become smaller and smaller as the prediction approaches $s_p$.
We will give an estimate of the error of the prediction $s_{PA}^{(N)}$ of the $P^N_1(s,s_0)$ considering the difference with respect to the previous PA, i.e., $P^{N-1}_1(s,s_0)$:
\bear
\label{eq:error}
\Delta s_N &\equiv & | s_{PA}^{(N)}\, \, -\,\, s_{PA}^{(N-1)}|\, .
\eear

\begin{figure}
  \includegraphics[width=2.25in]{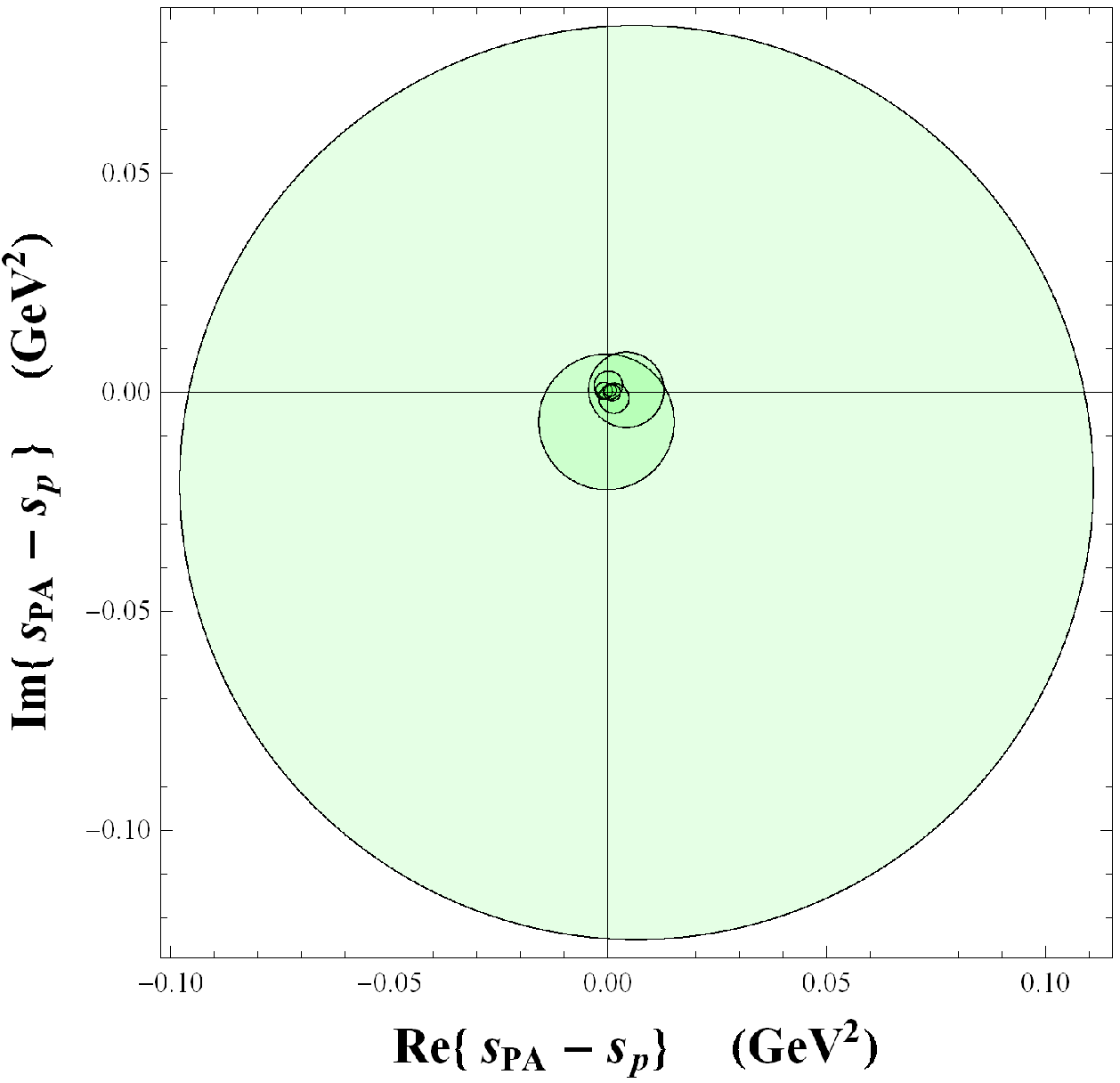}
  \includegraphics[width=2.25in]{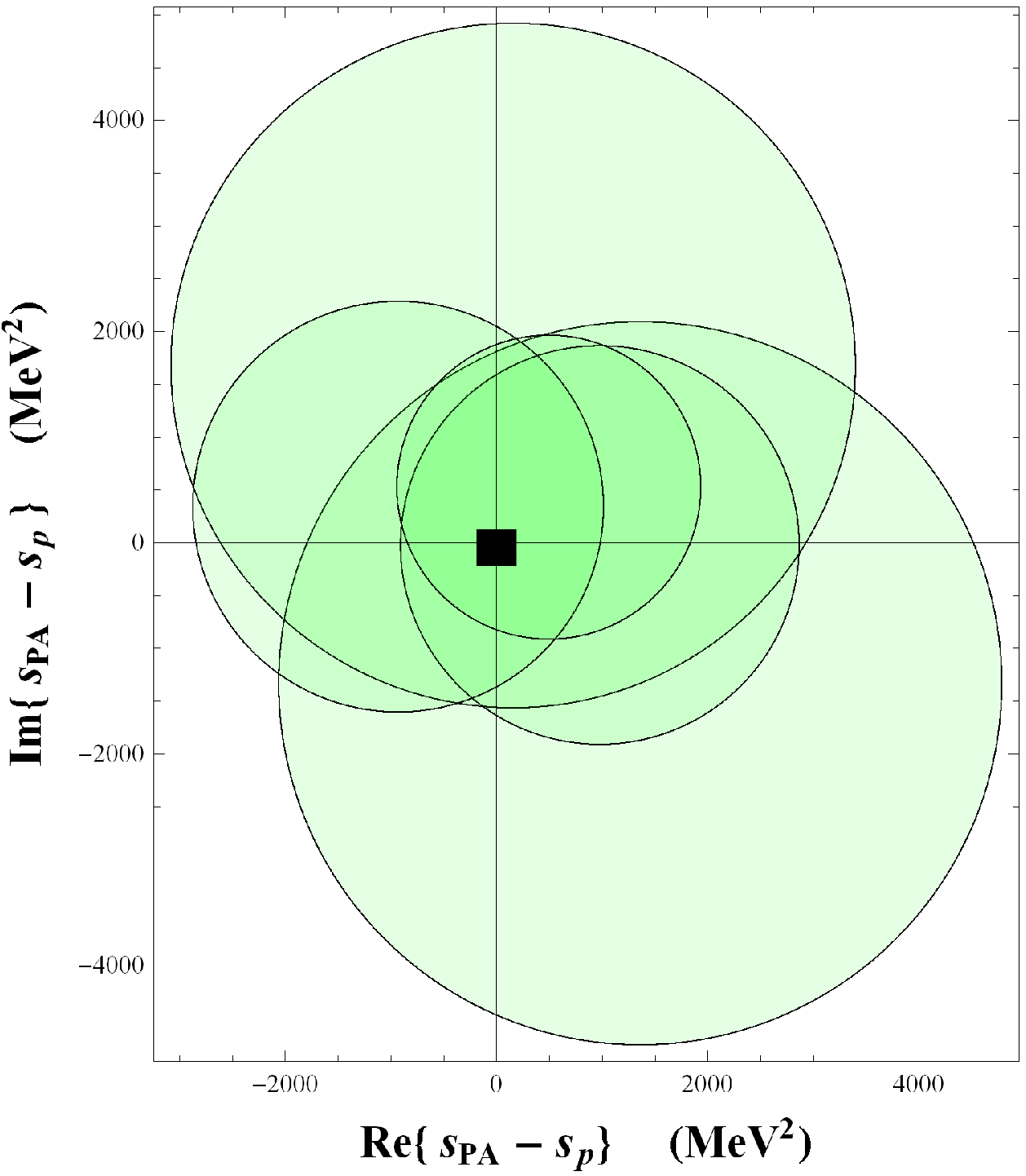}\\
  \caption{{\small
 Uncertainty regions for the genuine PA pole predictions in the case of a $\sigma$--like resonace
 with {\it model C}. {\bf Top:} results for $P^N_1(s,s_0)$ from $N=1$ up to $N=8$, in GeV$^2$. {\bf Bottom:} 
 same regions from $N=4$ up to $N=8$, in MeV$^2$.
  }}
  \label{fig.GPA-error}
\end{figure}

To illustrate how this error works, we will take the previous {\it model C} in terms of the $s$ variable
with a  $\sigma$-like resonance: PA center at $s_0=0.6$~GeV$^2$ and $s_p=(0.5- i 0.5/2)^2$~GeV$^2$.
In Fig.~\ref{fig.GPA-error} one can see the sequence of uncertainty regions.
For sake of clarity we actually provide $s_{PA}-s_p$,
in order to see deviations from the original pole $s_p$.
As the order $N$ of the $P^N_1(s,s_0)$ increases one obtains smaller and smaller circles. Even for such
wide resonance and inconvenient choice of variable as those considered in this example
($w$ would have shown a better convergence when $N\to\infty$),
it is easy to observe in Fig.~\ref{fig.GPA-error} that the $s_{PA}=s_p$ point is always contained in the error circles, which little by little converge to it.

Hence, we will consider this as our systematic error estimator and exemplify its use with a phenomenological
example. Moreover, we will study the dependence of the error size on the choice of the PA center $s_0$.
We will scan the possible $s_0$ points between the two thresholds $s_1^{th}$ and $s_2^{th}$
and optimize our PA determination by selecting  the $s_0^*$ point which minimizes the uncertainty
$\Delta s_N$.

\subsection{Fitting PA}

The second method we want to present consists, essentially, on fitting the experimental data with a PA sequence of higher and higher order. In such a way, one obtains information of the derivatives of $F(s)$ that will then be used to obtain the resonance pole and the residue positions as it is done in the previous subsections. In order to illustrate this fitting method with our physical {\it model C} in the s-variable, we generate a series of data points with zero error (pseudodata), which would represent an ideal experimental situation where all the uncertainty would be theoretical. We generate one-hundred points between the two-production thresholds (from $s=0$ to $s=1$) for both the modulus and the phase-shift of our model. This exercise should also prevent us against over-fitting problems.

Now,  instead of constructing the Pad\'e Approximants based on the derivatives around some energy point, we construct a sequence of  generic $P_1^N(s)$ and fit their various unknown parameters $a_n$ to these pseudodata. We fit it for the modulus and phase-shift of $F(s)$ and extract the optimal complex parameters $a_k$ for each  $P^N_1(s)$. Notice that this does not mean to fit $|F(s)|$ (or the VFF phase) with a $P^N_1(s)$. The modulus and phase-shift of the pseudodata are, respectively, fitted with the modulus and phase-shift of $P^N_1(s)$.

\begin{figure}
 \includegraphics[width=3.5in]{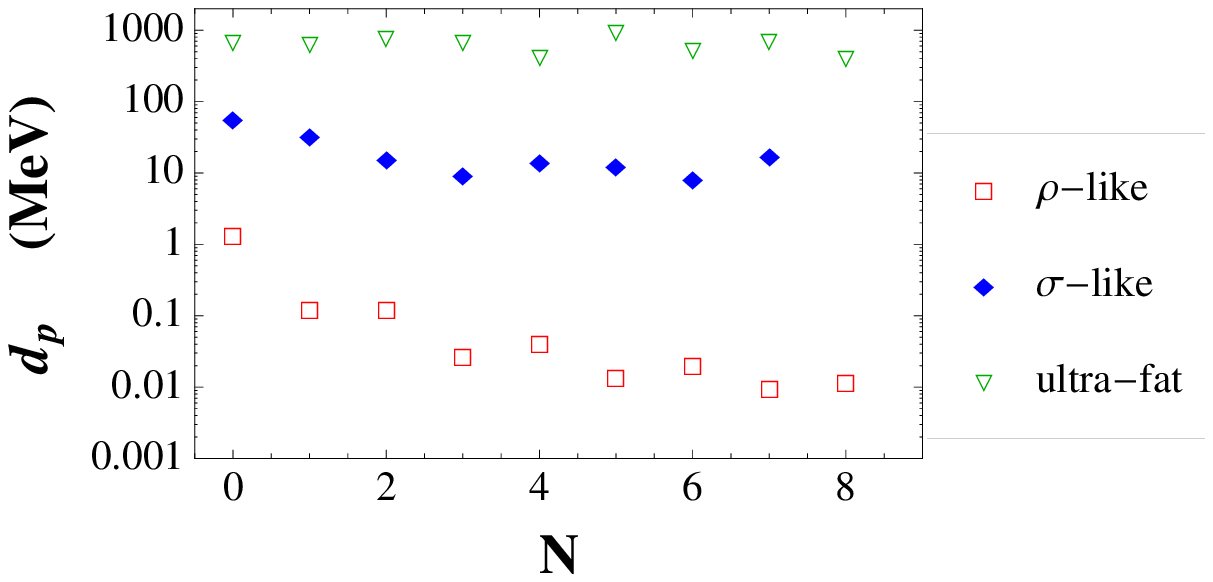}
 \includegraphics[width=3.5in]{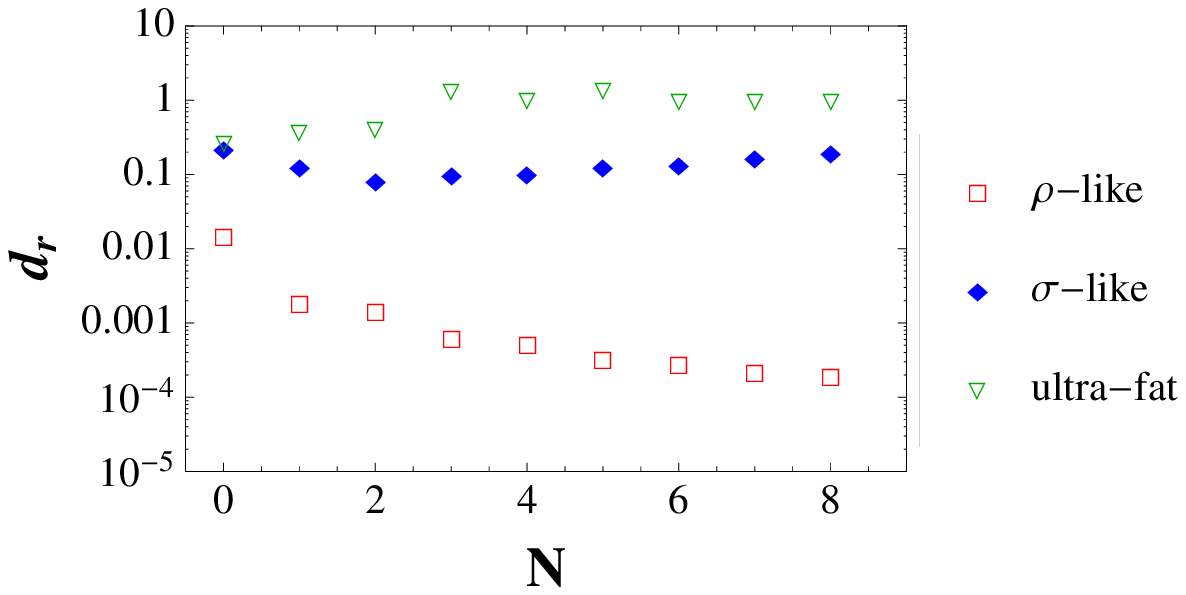}\\
  \caption{Rate of convergence of the $P^N_1(s)$ Pad\'{e} Approximant sequence
  corresponding to the {\it model C} in Eq.~(\ref{eq.models}). {\bf Top:} deviation from the
  true pole position in MeV (Eq.~(\ref{eq.dpole})).
{\bf  Bottom:} relative value of the deviations in the value of the residue (Eq.~(\ref{eq.dres})).
  For all the cases we used the PA center $s_0=0.6$~GeV$^2$. 
  }\label{fig:FPA}
\end{figure}

Once the $a_n$ parameters are known, we extract the position of the PA pole,  $s_{PA}$, as we did in the previous section, 
and compare it with the actual position thanks to the previous function $d_p$. The rate of convergence of our sequence for a $\rho$-like, for a $\sigma$-like, and for an {\it ultra-fat}-like resonance profiles are shown in Fig. \ref{fig:FPA}. The convergence is clear for the $\rho$-like resonance but, nonetheless, we observe that the convergence in the case of {\it fitting PAs} seems to be slightly  slower than in the previous  case with {\it genuine PAs}. For the $\sigma$-like resonance the convergence is slower than for the $\rho$-like resonance, and stabilizes at certain $N$ without further improvement. For an ultra-fat resonance, the convergence with the s-variable is not seen, neither for the resonance pole, nor for the residue. The results shown in Fig.~\ref{fig:FPA} can be improved upon (specially for the $\rho$-like and $\sigma$-like resonances) by enlarging the pseudodata set used.

\subsection{Comparison between different PA sequences}

\begin{figure}
\begin{center}
%
  \includegraphics[width=2.5in]{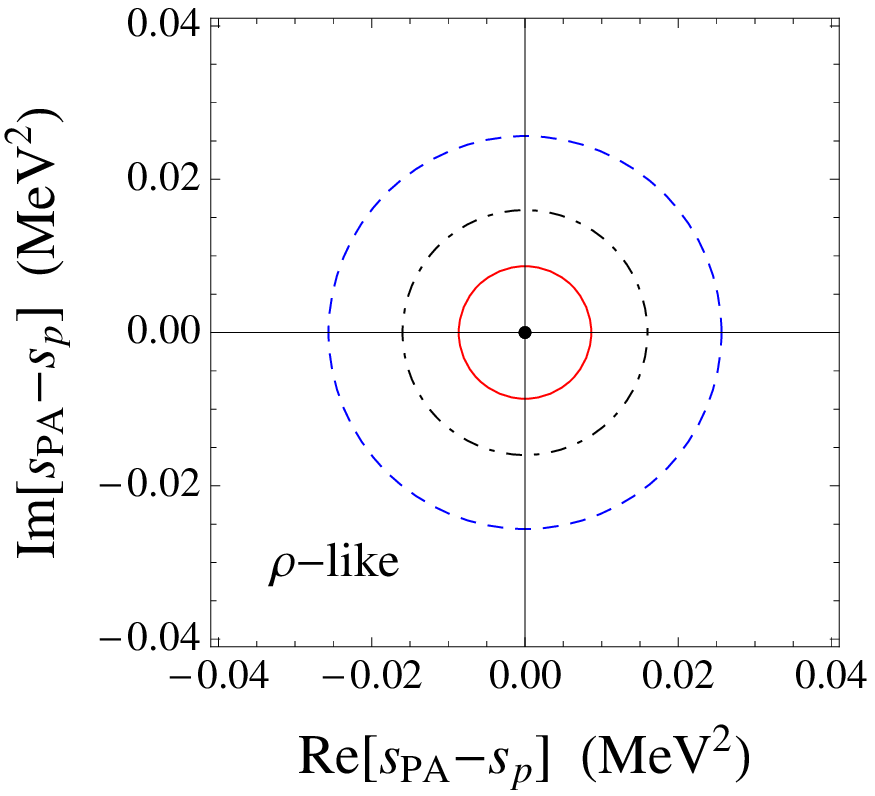}\\
    \vspace{0.5cm}
    \includegraphics[width=2.5in]{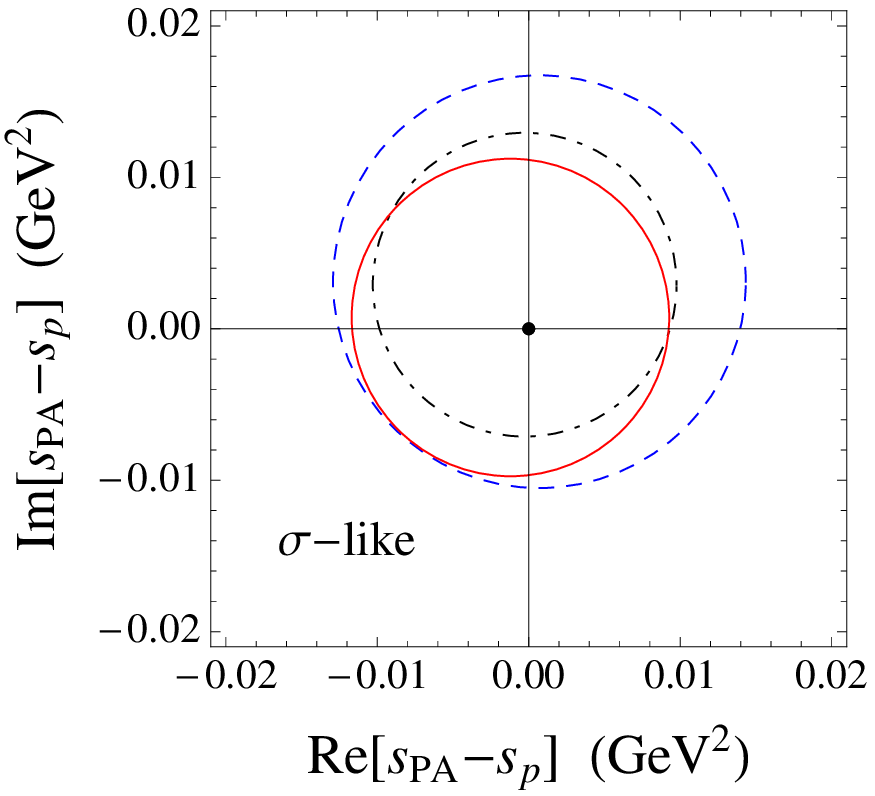}\\
  \caption{\small{
Comparison between the pole prediction using $P^0_4(s)$ (dashed blue), $P^3_1(s)$ (dotdashed black) and $P^2_2(s)$ (solid red) sequences for $\rho$-like (upper panel) and $\sigma$-like (lower panel) resonance profiles. Each circle represents the error ascribed to each approximant following the criteria $\Delta s$ defined in Eq.~(\ref{eq:error}).}}\label{fig:N6}
\end{center}
\end{figure}

In this subsection we briefly comment on different PA sequences that one may consider for locating resonance poles when only a finite number of derivatives of the function $F(s)$ to be approximated are given. Montessus' theorem states convergence for a $P^N_1(s,s_0)$ sequence  in case there is only one resonance pole in our disk of convergence. Pad\'e Theory, however, provides with other convergence theorems that would be appropriate for locating resonance poles in this situation, such as Pommerenke's theorem~\cite{Pommerenke,Baker,Masjuan:2007ay} which states convergence to a meromorphic function for a $P^{N+J}_N(s,s_0)$ sequence with $J\leq -1$, and $N\rightarrow \infty$. Likewise, the so-called diagonal sequence ($J=0$), would work for a single-resonance pole~\cite{Pommerenke}. On the other hand, a sequence $P^0_N(s,s_0)$ has been commonly used in the literature~\cite{Caprini:2008fc,Martin:2011cn} to locate the resonance pole on the 2RS of the $\pi \pi $ S wave, looking for the $\sigma$ of $f_0(500)$ meson. However, we do not know of any convergence theorem for this kind of sequences.

In Fig.~\ref{fig:N6} we compare the three sequences considering the particular case that only the value of the function and its first four derivatives for the {\it model C} are known, for both a $\rho$-like and a $\sigma$-like resonance profiles. We optimized the energy point where to center our approximants. For the first case we use $s_0=0.6$ and for the second $s_0=0.4$. Five inputs $a_n$ imply that we can construct $P^3_1(s,s_0)$, $P^2_2(s,s_0)$ and $P^0_4(s,s_0)$. The circles represent the error ascribed to each PA for each approximant following the criteria $\Delta s$ defined in Eq.~(\ref{eq:error}).

For the $\rho$-like resonance profile the hierarchy of predictions is clear: the $P^N_N(s,s_0)$ is the best choice and the $P^N_1(s,s_0)$ goes behind. The $P^0_N(s,s_0)$ is the worse scenario. Notice that the errors are in MeV$^2$. The situation is different for a $\sigma$-like resonance, where both $P^N_N(s,s_0)$ and $P^N_1(s,s_0)$ provide similar predictions but again the $P^0_N(s,s_0)$ represents the worse scenario. In this second case, the errors are in GeV$^2$.

Notice, however, that the $P^N_N(s,s_0)$ sequence ``grows" in steps of two-by-two derivatives and that the error shown in Fig.~\ref{fig:N6} is defined as the difference of the pole predictions from the $P^2_2(s,s_0)$ and the $P^1_1(s,s_0)$.

\section{Phenomenological examples}\label{sec:pheno}

In the previous section we investigate the role of PAs as an appropriate tool to determine resonance parameters. We considered three analytical models analyzed in two different fashions. In this section we study phenomenological examples following the same way: we apply the ``genuine PA" for searching the $\kappa$ pole and the ``fit PA" for the $\rho$ one.

\subsection{Genuine PA:
$K\pi$ scattering    and the $\kappa$  pole}

We will make use of the outcome from the Roy-Steiner equation for the
$K\pi$ scattering amplitude $T(s)^{I=1/2}_{J=0}$~\cite{Moussallam:2004}.
Here we will just exemplify our ``genuine PA" method and will not go deeper into the statistical error analysis,
which should be also properly accounted in the final determination of the resonance pole.

In the elastic region, the partial-wave scattering amplitude shows the form
\be\label{eq:scat}
\sigma(s) T^{1/2}_{0}(s) \,\,\,=\,\,\,
e^{i\delta^{1/2}_{0}(s)}\, \sin\delta^{1/2}_{0}(s)\,\,\,=\,\,\, \Frac{1}{\hat\psi(s)\, -\, i}\, ,
\ee
with the phase-space factor $\sigma(s)=\lambda^\frac12 (s,m_\pi^2,m_K^2)/s = 2 k(s)/\sqrt{s}$
and the analytical extension $\hat\psi(s)=\cot\delta^{1/2}_{0}(s)$ to the complex $s$--plane. Notice that Eq.~(\ref{eq:scat}) is implicitly evaluated at $s+i\epsilon$  
in the 1RS. We will use the PA to compute its analytical extension to the 2RS. 

\begin{figure}
  \includegraphics[width=7cm]{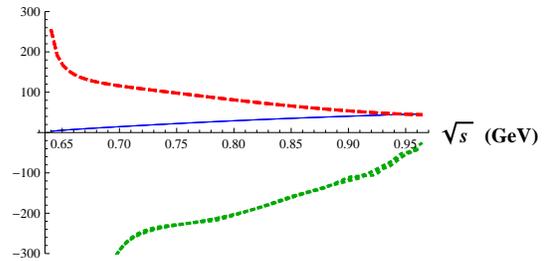}
  \caption{
  {\small Partial-wave phase-shift $\delta^{1/2}_{0}(s)$ (solid blue),
                   its first derivative $\frac{d}{ds}\delta^{1/2}_{0}(s)$  (dashed red)
                   and second derivative
                   $\frac{d^2}{ds^2}\delta^{1/2}_{0}(s)$ (dotted green)~\cite{Moussallam:2004},
                   in units of {\it degree}, {\it degree} GeV$^{-2}$ and {\it degree} GeV$^{-4}$,
                   respectively.}
 }\label{fig.phaseKpi}
\end{figure}

With the phase-shift $\delta(s)$ and its subsequent derivatives $\frac{d^n\delta}{ds^n}$
at a given point $s_0\in (s_1^{th}\, ,\, s_2^{th})$ between the lowest and second threshold, $s_1^{th}=(m_\pi+m_K)^2$ and  $s_2^{th}=(m_K+m_\eta)^2$, respectively (if multiparticle channels are neglected), we may construct a $P^N_1(s,s_0)$  of the partial-wave amplitude $T(s)$ around $s_0$,
with the series coefficients  $a_k=\frac{1}{k!} \frac{d^kT}{ds^k}\bigg|_{s=s_0}$.
The experimental value of the firsts derivatives of the phase-shift are shown in
Fig.~\ref{fig.phaseKpi}~\footnote{
We thank B. Moussallam for his help with the $K\pi$
phase-shift from Ref.~\cite{Moussallam:2004} and its derivatives.}. We will employ the phase-shift derivative up to third order, which implies a $P^N_1(s,s_0)$ sequence up to $N=2$.

\begin{figure}
\begin{center}
  \includegraphics[width=8cm]{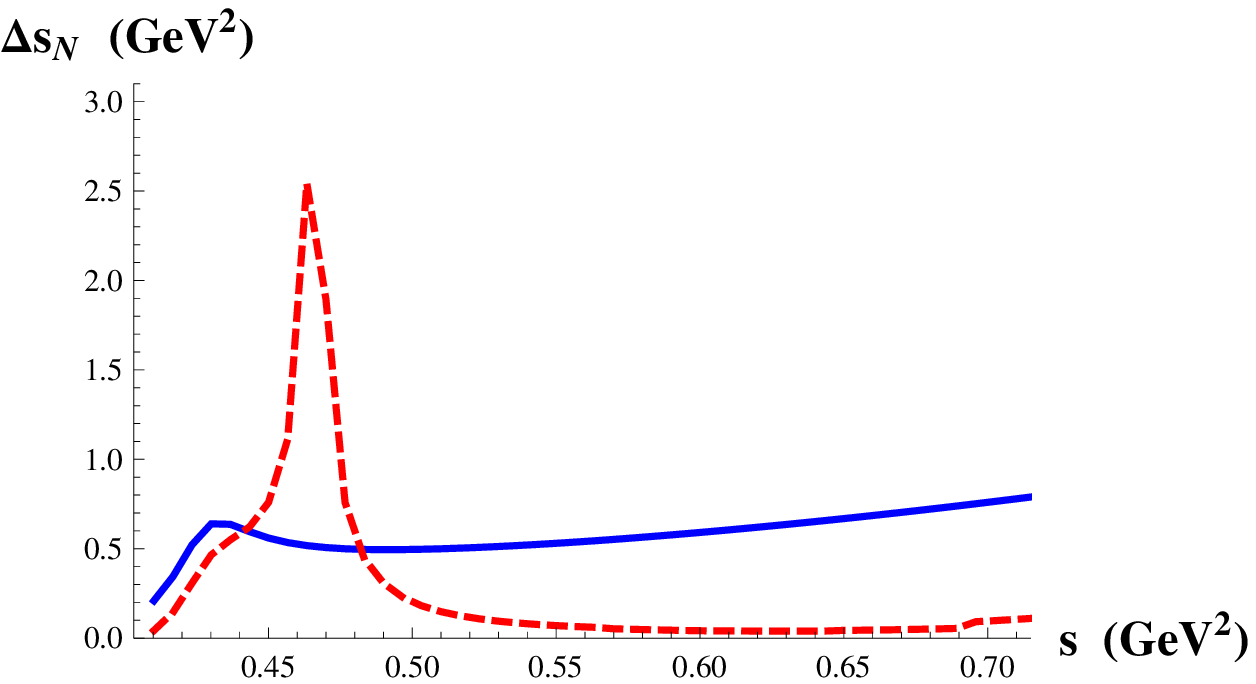}
  \includegraphics[width=6cm]{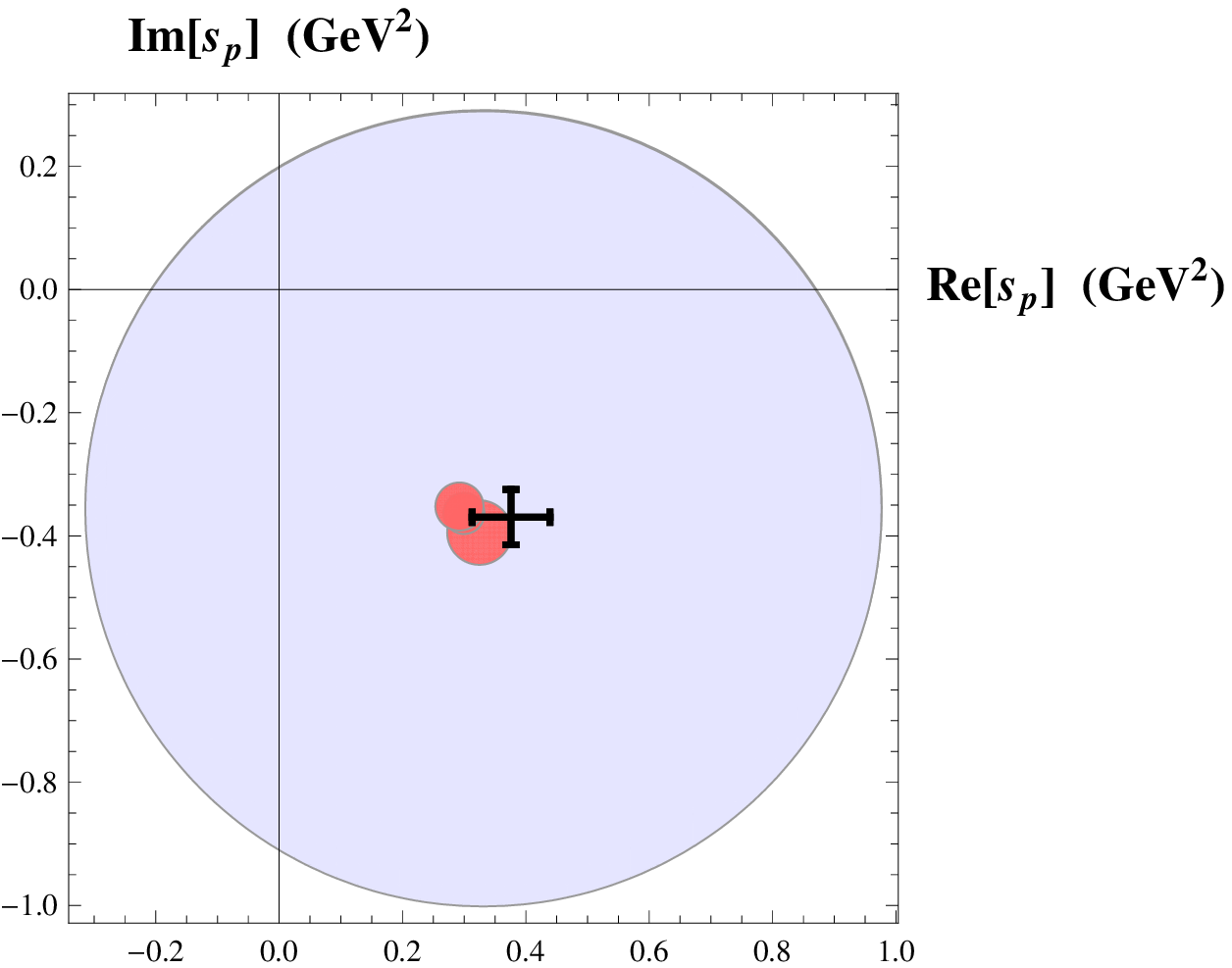}
  \caption{\small{
  {\bf Top:} Distance between prediction for $\Delta s_1$ (solid blue) and $\Delta s_2$ (dashed red).
 {\bf  Bottom:} Pole determinations with the corresponding associated error from the $P^1_1$ (light blue) and
  and the $P^2_1$ approximant (darker red) compared to previous determinations
  (black error bar)~\cite{PDG2012}.
  We have plotted together the outcome from the $K\pi$ phase-shift for three different interpolations~\cite{Moussallam:2004}. The obtained error regions overlap and the predictions are found
  to be very stable.
  }}\label{fig.dist-kappa-predi}
\end{center}
\end{figure}

One can see  in Fig.~\ref{fig.dist-kappa-predi}
that in general the distance $\Delta s_N$ between the $P^{N-1}_1(s,s_0)$ and $P^N_1(s,s_0)$ pole predictions, decreases as the order of the PA  increases (Fig.~\ref{fig.dist-kappa-predi}, upper panel).   In our case
--where we extracted the PA up to  $N=2$--, one can observe that
only when the PA center $s_0$ is close to the $K\pi$ threshold one finds $\Delta s_2\gsim \Delta s_1$.
We focus on the $s_0$ range where $\Delta s_2 < \Delta s_1$ and then take
the prediction with minimal error $\Delta s_2$ as our best determination.
The optimal PA  center is found  at
$s_0\sim 0.6$~GeV$^2$. As we said, we discard the solutions with $s_0<0.45$~GeV$^2$ due to its
proximity to the $K\pi$ branch cut singularity.
In Fig.~\ref{fig.dist-kappa-predi}, lower panel, we show our predictions for $P^N_1(s,s_0)$ with $N=1$ and $N=2$ with its corresponding $\Delta s_N$ error. They
are depicted  by the shaded uncertainty regions provided
in Fig.~\ref{fig.dist-kappa-predi} --bottom--. One can see that a  clear convergence to
previous phenomenological determinations is obtained~\cite{Moussallam:2004,PDG2012}.

This shows that it is possible to use PAs for the determination of resonance poles
and the study of this kind of amplitudes. Indeed, with a relatively small amount of information
($\delta(s_0)$ and its first three derivatives) we have achieved a precision and accuracy similar  to
that from alternative procedures, which are in general more involved~\footnote{Notice, however, that the fair determinations of the input derivatives originally stem from the Roy-Steiner analysis~\cite{Moussallam:2004}}.
This kind of techniques may be quite useful in the cases when for some reason some data can  be very
well determined in a local range of energy, as it is based on the value of the first derivatives, providing
a wider range of validity and convergence than simple Taylor series.

Nevertheless, no detailed analysis of the errors in the $\delta(s_0)$ derivative inputs is performed
in this work and hence the results presented in Fig.~\ref{fig.dist-kappa-predi} are just a first estimate.
The inclusion of the statistical uncertainties would certainly enlarge the error of our final determination.

\subsection{PA as fitting functions: the $\rho(770)$  pole determination}

\begin{figure}[!h]
\begin{center}
  \includegraphics[width=6.5cm]{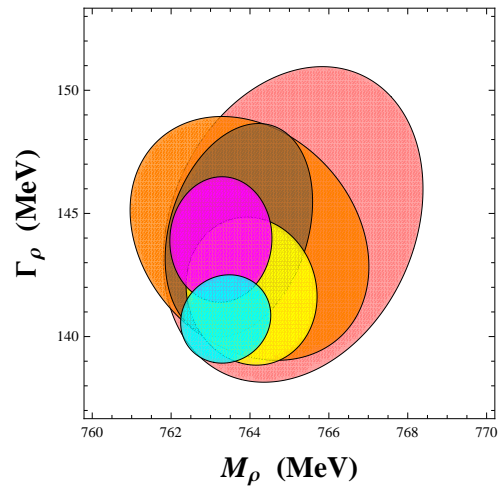}
\caption{{\small
68\% CL regions for the rho pole mass and width from
the different $P^N_1$ fits.
The smallest (cyan) ellipse provides the prediction for $N=3$ and
the following growing
orders in $N$  are given by the ovals with larger and larger size.
  }}
\label{fig.CL}
\end{center}
\end{figure}

We proceed now to analyze the final compilation of ALEPH $\pi\pi$-VFF data for the squared modulus
$|F_{\pi\pi}(q^2)|^2$~\cite{ALEPH,Davier:2005xq}
and the $I=J=1$ $\pi\pi$ scattering phase-shift $\delta_1^1(s)$,
identical to the $\pi\pi$-VFF phase-shift in
the elastic  region $4m_\pi^2<q^2<4 m_K^2$
(if multipion channels are neglected).
This will be the range of application of $P^N_1(s)$ analysis.
For $N\geq 3$ the fit $\chi^2$ already lies within
the 68\% confidence level (CL)  and becomes statistically acceptable.
Their corresponding 68\% CL regions for the pole mass and width predictions
are shown in Fig.~\ref{fig.CL}.
The regions from the different fits  overlap each other
in a compatible way.  The allowed ranges become larger and
larger as $N$ grows and the fit contains more and more free parameters.

At this point one needs to reach a compromise. On one hand the
experimental (fit) errors have an statistical origin and increase
as one considers higher order $P^N_1(s)$,
with a larger number of parameters.  On the other,  the
systematic theoretical (Pad\'e Approximant)  error decreases as $N$  increases and the PA converges to the actual VFF.
In the present work we have taken $N=6$ as our  best
estimate as the new parameters of $P^N_1(s)$ with $N\geq 7$
turn out to be all
compatible with zero, introducing no information with respect to $P^6_1(s)$.
Furthermore,  the different models studied before show  that in any case
the theoretical errors for mass and width
result smaller than $10^{-1}$--$10^{-2}$~MeV  for $N\geq 6$, being negligible compared to
the $\cO(1$~MeV$)$  experimental errors, see Figs.~\ref{fig:FPA} and \ref{fig:N6}.
This yields the determinations
\begin{equation}\label{eq:rhoresult}
M_\rho\,=\, 763.7\pm 1.2\, \mbox{MeV}\,,
\qquad
\Gamma_\rho\,=\, 144\pm 3 \,\mbox{MeV}\, ,
\end{equation}
which is found in reasonable agreement with former
determinations obtained from more elaborated procedures
and  with similar size for the uncertainties, collected in Table~\ref{tab:rho}.

\begin{table}[htdp]
\caption{Collection of different predictions for the $\rho$ meson resonance parameters compared to our result in Eq.~\ref{eq:rhoresult}.}
\begin{center}
\begin{tabular}{|c|c|c|}
\hline
& $M_{\rho}$ (MeV) & $   \Gamma_{\rho}$ (MeV)\\
\hline
\mbox{[Ananthanarayan {\it et al.}~\cite{Roy} ]} &  $762.5\pm 2 $ & $142\pm 7$\\
\mbox{[IAM~\cite{IAM-pole-pred,IAM-pole-pred2} ]} & $754\pm 18$ &  $148 \pm 20$ \\
\mbox{[Zhou {\it et al.}~\cite{rho-Zheng} ]} & $763.0\pm 0.2$ & $139.0\pm 0.5$ \\
\mbox{[Pich and SC~\cite{Cillero-VFF} ]} & $764.1\pm 2.7^{+4.0}_{-2.5}$ & $148.2\pm 1.9^{+1.7}_{-5.9}$\\
\mbox{[Dumm and Roig~\cite{Dumm:2013zh} ]} & $744.1 \, $--$\,  761.1$ & $142.7 \,$--$\, 149.9$\\
\hline
\mbox{[This work]} & $763.7\pm 1.2$ & $144\pm3$\\
\hline
\end{tabular}
\end{center}
\label{tab:rho}
\end{table}

\section{Pad\'e approximants and Breit-Wigner mass}\label{sec:comparison}

In the case of an elastic  two-particle channel
the partial-wave scattering amplitude shows the form
\begin{eqnarray}
\sigma(s) T(s) &=&  \sin\delta(s) \,e^{i\delta(s)}\,
=\, \Frac{1}{\cot\delta(s)\, -\, i}\, ,
\end{eqnarray}
where the complex amplitude is parametrized by just one real parameter,
the phase-shift $\delta(s)$.

As we have seen, the $P^N_1(s,s_0)$ sequence is very well suited for the search
of a possible resonance pole in the 2RS in this kind of scenarios.
Of particular interest is the first PA of the series:
\bear
P^0_1(s,s_0) &=& \Frac{a_0}{1-\Frac{a_1}{a_0}(s-s_0)}\, =\, \Frac{(a_0)^2/a_1}{(s_0+a_0/a_1)\, -\, s} \, .
\eear
If we place the PA center at the Breit-Wigner mass ($s_0=M_{BW}^2$) where $\delta(M_{BW}^2)=\pi/2$,
then one finds the Taylor coefficients $a_0=i$ and $a_1= - \delta'(M_{BW}^2) = - (M_{BW}\Gamma_{BW})^{-1}$.
We have used the Breit-Wigner width $\Gamma_{BW}$ definition in the last relation.
This allows us to rewrite the PA in the familiar form
\begin{eqnarray}
P^0_1(s,M_{BW}^2) &=& \Frac{M_{BW} \Gamma_{BW}}{M_{BW}^2-s-i M_{BW} \Gamma_{BW}} \, .
\end{eqnarray}
Hence, the simplest PA readily provides a prediction for the resonance pole and residue (see Eq.~(\ref{eq:poleres})):
\bear
s_{PA}^{(0)} &=& s_0\, +\, a_0/a_1
\nn\\
&= & \, M_{BW}^2 \, -\, \Frac{i}{\delta'(M_{BW}^2)}
\nn\\
&=&  M_{BW}^2 \, -\, i\, M_{BW}\Gamma_{BW}\, ,
\nn\\
\label{eq.BW-P01}
\eear
and 
\bear
Z_{PA}^{(0)} =-a_0^2/a_1= - M_{BW}\Gamma_{BW}\, .
\eear

If we construct the next approximant, the $P^1_1(s,M_{BW}^2)$, one obtains the pole prediction
\bear
s_{PA}^{(1)} &=& s_0\, +\, a_1/a_2\\
&= & \, M_{BW}^2 \, -\, \Frac{i}{\delta'(M_{BW}^2)} \,
\bigg(\,1 \, -\, \Frac{i\delta ''(M_{BW}^2)}{2(\delta'(M_{BW}^2))^2  }  \bigg)^{-1}
\nn\\
&=&  M_{BW}^2   -  i  M_{BW}\Gamma_{BW}
\nn\\
&&\qquad\qquad\times
\bigg( 1  - \Frac{i}{2} M_{BW}^2 \Gamma_{BW}^2 \delta ''(M_{BW}^2) \bigg)^{-1} \, .\nn
\nn\\\nn
\label{eq.BW-P11}
\eear

Its corresponding residue reads
\bear
Z_{PA}^{(1)} &=&-a_1^3/a_2^2 \nn\\
&=& -\, \Frac{1}{\delta'(M_{BW}^2)} \,
\bigg(\,1 \, -\, \Frac{i\delta ''(M_{BW}^2)}{2(\delta'(M_{BW}^2))^2  }  \bigg)^{-2}\\
&=&   - M_{BW}\Gamma_{BW}\times
\bigg( 1  - \Frac{i}{2} M_{BW}^2 \Gamma_{BW}^2 \delta ''(M_{BW}^2) \bigg)^{-2} \, .\nn\\\nn
\eear\nn

It is not difficult to show that the expressions~(\ref{eq.BW-P01} --\ref{eq.BW-P11})
are in agreement  with those in Ref.~\cite{Nieves:2009kh}. Based on the arguments therein
one can easily prove that in meson--meson scattering,  if $s_p$ scales like $\cO(N_C^0)$
at large $N_C$,  then one has
\bear
s_p=s_{PA}^{(0)}\,+\, \cO(N_C^{-2}) \,=\, s_{PA}^{(1)}\, +\, \cO(N_C^{-4})\, ,
\eear
for the PA center $s_0=M_{BW}^2$, with corrections always smaller than the one provided by the half-width rule~\cite{Masjuan:2012gc,Masjuan:2012sk,Masjuan:2013xta}.

Phenomenologically, these predictions were found to be fairly good even for broad resonances such as the $\sigma$ meson, reaching uncertainties of the order of 10\% or smaller in the determination of the pole mass and width~\cite{Nieves:2009kh}.

\section{Conclusions}\label{sec:conclusions}

In this article we have provided a simple procedure for the extraction of resonance poles
in a theoretically sound way.  We rely on general mathematical theorems that ensure the convergence of some given sequences of rational approximants, in particular Pad\'e Approximants, as far as one remains within their range of applicability. For the Montessus' theorem this means
that the resonance pole must be contained in the maximally large disk contained between the two thresholds
$s_1^{th}$ and $s_2^{th}$. Moreover, instead of working with the $s$--variable,
one may improve or even ensure the convergence of the sequence
by means of convenient variable transformations
(like the conformal mapping $w(s)$).
Likewise, we have provided a reliable estimate of the systematic uncertainty of our PA estimates.

Finally, after illustrating the techniques with some theoretical models,
we have applied our procedure to real phenomenology. Through the ``genuine PA" approach (approximants built from the derivatives of the function to be approximated) we have shown
how the $P^2_1(s)$ to the $K\pi$ scattering amplitude $T^{1/2}_0(s)$
provides already an excellent prediction of the broad $\kappa$ resonance pole.
We have also employed the PA as fitting functions to extract the $\rho(770)$ pole position from
the experimental $\pi\pi$ vector form-factor and $\pi\pi$ scattering phase-shift $\delta^1_1(s)$,
in very good agreement with alternative determinations and with similar error size, accounting for systematic errors coming from the procedure itself.

Exploiting the analyticity of the matrix elements, this procedure can be easily applied to a vast
set of observables: the $\sigma$ pole determination in $IJ=00$ $\pi\pi$ scattering;
heavy quark resonances in $e^+e^-\to D \overline{D}$ below the $D\overline{D}^*$ threshold;
precise determination of the $\Delta(1232)$ in $\pi N$--scattering, search for exotics, Dalitz decay parameterizations, etc.

Further analysis in these directions are relegated to future works.

\section*{Acknowledgements:}

We would like to thank L. Tiator and B. Moussallam for comments on the manuscript. B. Moussallam is also acknowledged for his help  with the $K\pi$ phase-shift from Ref.~\cite{Moussallam:2004}.  S.C. would like to thank the University of Mainz for its hospitality.
This work has been partially supported by the MICINN, Spain, under contract FPA2010-17747 andÊ Consolider-Ingenio CPAN CSD2007-00042, by the Italian Miur PRIN 2009, the Universidad CEU Cardenal Herrera grant PRCEUUCH35/11, the MICINN-INFN fund AIC-D-2011-0818 and by the Deutsche Forschungsgemeinschaft DFG through the Collaborative Research Center ``The Low-Energy Frontier of the Standard Model" (SFB 1044). We thank as well the Comunidad de Madrid through Proyecto HEPHACOS S2009/ESP-1473 and the Spanish MINECO Centro de excelencia Severo Ochoa Program under grant SEV-2012-0249.

\end{document}